\documentclass[12pt]{article}
\usepackage{graphicx}
\usepackage{amsmath}
\usepackage{amssymb}
\usepackage{caption2}
\begin{document}
\newcommand  {\ba} {\begin{eqnarray}}
\newcommand  {\ea} {\end{eqnarray}}
\renewcommand{\thefootnote}{\fnsymbol{footnote}}
\renewcommand{\figurename}{FIG.}
\renewcommand{\captionlabeldelim}{.~}

\begin{flushright}
 {\large\bf BIHEP-TH--2002-54}
\end{flushright}
\vskip 3.0cm

\begin{center}
{\Large\bf \emph{CP} violation in chargino decays in the  MSSM}
 \vskip 1.0cm
{\bf Wei Min Yang\footnote{Email address: yangwm@mail.ihep.ac.cn} and Dong Sheng Du}\\
 {\em CCAST(World Laboratory), P.O.Box 8730, Beijing 100080, China
 and Institute of High Energy Physics, P.O.Box 918(4), Beijing 100039, China}
 \vskip 2.0cm
\textbf{ABSTRACT}
\end{center}

In the minimal supersymmetric standard model (MSSM) with complex parameters, supersymmetric loop effects can lead
to \emph{CP} violation. We calculate the rate asymmetries of decays of charginos into the lightest neutralino and
a $W$ boson on the basis of the most important loop contributions in the third generation squark sectors. It turns
out that the \emph{CP} violating asymmetries can be a few per cent in typical regions of the parameter space of
the MSSM. These processes would provide very promising channels for probing \emph{CP} violation in the MSSM at
future high-energy colliders.\\


\newpage

\begin{center}
\textbf{I. INTRODUCTION}
\end{center}

Searching for new \emph{CP} violating effects beyond the standard model (SM), is a important and interesting work
for theoretical and experimental high-energy physicist. The MSSM are currently considered as the most
theoretically well motivated extensions of the (SM) \cite{1,2}. In comparison with the sole CKM phase in the SM,
it contains more diverse sources of \emph{CP} violation through complex soft-SUSY-breaking parameters \cite{3a}.
These new soft phases affect not only \emph{CP} violating observables but can have significant impacts in a
variety of places \cite{3}, including $g_{\mu}-2$, electric dipole moments (EDMs), \emph{CP} violation in the $K$
and $B$ systems, the baryon asymmetry of the universe, cold dark matter, superpartner production cross sections
and branching ratios, and rare decays. Although there are some experiments that suggest some of the phases are
small, mainly the neutron and electron EDMs \cite{4}, it has recently been realized that \emph{CP} violating
phases associated with the third generation trilinear soft-breaking terms might be large and can induce sizable
\emph{CP} violation in the Higgs and the third generation sfermion sectors through loop corrections \cite{5,6}.
Such phases may allow baryogenesis and do not necessarily violate the stringent bound from the nonobservation of
(EDMs) \cite{7}. In fact, some of these phases can be $\mathcal{O}(1)$, so as to provide non-SM sources of
\emph{CP} violation required for dynamical generation of the baryon asymmetry of the universe \cite{8a,8b}. The
important implication of the \emph{CP} violating phases in the search for supersymmetry has received growing
attention.

The decays of charginos $\widetilde{\chi}_{i}^{+}$ $(i=1,2)$ and neutralinos $\widetilde{\chi}_{l}^{0}$ $(l=1-4)$
in the MSSM under \emph{CP} conservation have been extensively studied \cite{9}. For most of the parameter space
of the MSSM, the decays of the heaviest chargino $\widetilde{\chi}_{2}^{+}$ and the two heavier neutralinos
$\widetilde{\chi}_{l}^{0}$ $(l=3,4)$ will be dominated by two-body tree-level processes in which the final states
include two possible classes: (i) a lighter neurralino or chargino plus a $W$, $Z$, or Higgs boson; and (ii)
channels involving squarks or sleptons if they are kinematically allowed. For the lightest chargino
$\widetilde{\chi}_{1}^{+}$ and the next lightest neutralino $\widetilde{\chi}_{2}^{0}$, the mass difference
between them and the lightest neutralino $\widetilde{\chi}_{1}^{0}$ is not usually so large, therefore their
dominant decay modes are three-body tree-level decays through a virtual vector boson or Higgs boson mediation. But
in some parameter region, the mass splitting between them can exceed the mass of a vector boson or Higgs boson, in
this case, the type (i) will be the most important decay modes of them. Furthermore, if they are enough heavy, the
decay channels of the type (ii) are also possible. In fact, the above decays can also occur at one-loop level via
the final state interactions if they are kinematically accessible, furthermore, the \emph{CP} violating phases can
directly affect the couplings of charginos and neutralinos to the third generation sfermions. Therefore, the
interferences of the tree diagrams of these decays with the corresponding one-loop diagrams can make their partial
decay widths difference from that of corresponding \emph{CP} conjugate processes. In this paper, as an example, we
will focus on \emph{CP} violation in the decays of charginos into the lightest neutralino and a $W$ boson as the
two-body-decay branching ratios can be large. It can be measured by the rate asymmetry:
 \ba
 A_{cp}=\frac{\Gamma(\widetilde{\chi}_{i}^{+}\rightarrow\widetilde{\chi}_{1}^{0}W^{+})-
 \Gamma(\widetilde{\chi}_{i}^{-}\rightarrow\widetilde{\chi}_{1}^{0}W^{-})}
 {\Gamma(\widetilde{\chi}_{i}^{+}\rightarrow\widetilde{\chi}_{1}^{0}W^{+})+
 \Gamma(\widetilde{\chi}_{i}^{-}\rightarrow\widetilde{\chi}_{1}^{0}W^{-})} \,.
 \ea
Of course, analogous asymmetries can also be discussed for the other decay channels of charginos and neutralinos.

The remainder of this paper is organized as follows: in Section II we list the relevant couplings and give the
analytical formulas of the decay rate asymmetries. In Sec.III, we present detailed numerical results. Our
conclusions are summarized in Sec.IV. Appendix A outlines the necessary masses and mixing matrices, and appendix B
contains the expressions of the form factors.

\begin{center}
\textbf{II. RELEVANT COUPLINGS AND DECAY RATE ASYMMETRY}
\end{center}

In order to calculate the rate asymmetry of decays of charginos into the lightest neutralino and a $W$ boson, in
the appendix A we review briefly the masses and mixing of the chargino, neutralino and sfermion sectors of the
MSSM. As mentioned in last section, although the tree-level decays of charginos is invariant under \emph{CP}, a
nonvanishing value of the rate asymmetry $A_{cp}$ in Eq.~(1) is generated at one-loop level by complex couplings.
Among others, the most significant \emph{CP} violating effects arise from the trilinear couplings of the third
generation $A_{t,b,\tau}$ \cite{10}, especially from squark sectors because of the Yukawa characteristic factor
$m_{q}$ ($m_{q}$ is the mass of quark $q$) in the chargino/neutralino-quark-squark couplings. Therefore, one can
expect that the most important contributions come from one-loop quark-squark exchange diagrams, as shown in
Fig.~1~(a-d). The relevant interaction Lagrangian are listed as follows \cite{1}:
\begin{figure}[t]
\centering
\includegraphics[totalheight=7.5cm]{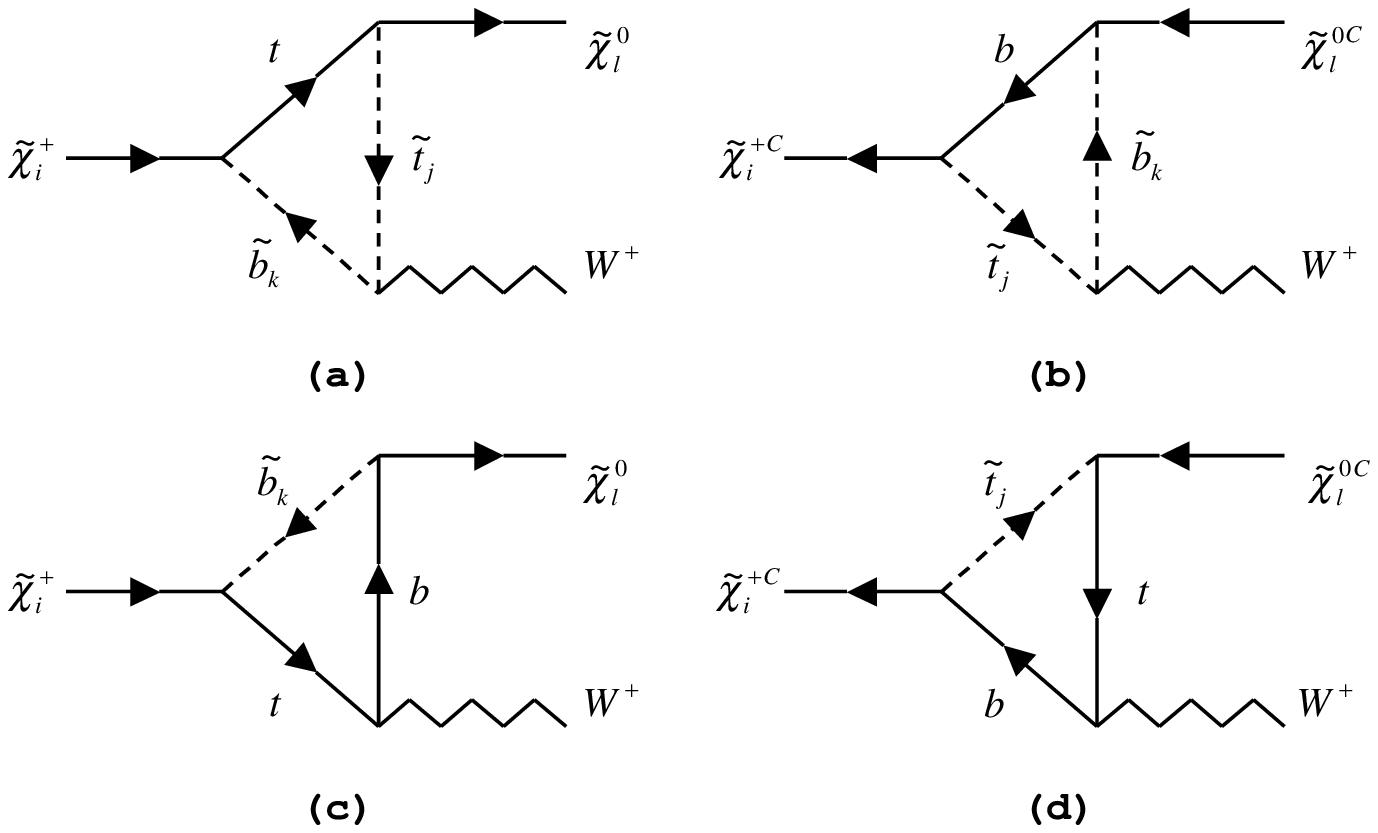}
\caption{The relevant one-loop diagrams for the decay
$\widetilde{\chi}_{i}^{+}\rightarrow\widetilde{\chi}_{1}^{0}W^{+}$.}
\end{figure}

\begin{alignat}{1}
\mathcal{L}_{\widetilde{\chi}^{+}q\widetilde{q}^{'}}
 &= \overline{t}(A^{L}_{ik}P_{L}+A^{R}_{ik}P_{R})\widetilde{\chi}_{i}^{+}\widetilde{b}_{k}
     +\overline{b}(B^{L}_{ij}P_{L}+B^{R}_{ij}P_{R})\widetilde{\chi}_{i}^{+c}\widetilde{t}_{j} \nonumber\\
 &\ \ \ +\overline{\widetilde{\chi}_{i}^{+}}(A^{R*}_{ik}P_{L}+A^{L*}_{ik}P_{R})t\widetilde{b}_{k}^{*}
     +\overline{\widetilde{\chi}_{i}^{+c}}(B^{R*}_{ij}P_{L}+B^{L*}_{ij}P_{R})b\widetilde{t}_{j}^{*} \,,\nonumber\\
\mathcal{L}_{\widetilde{\chi}^{0}q\widetilde{q}}
 &= \overline{t}(C^{L}_{lj}P_{L}+C^{R}_{lj}P_{R})\widetilde{\chi}_{l}^{0}\widetilde{t}_{j}
     +\overline{b}(D^{L}_{lk}P_{L}+D^{R}_{lk}P_{R})\widetilde{\chi}_{l}^{0}\widetilde{b}_{k} \nonumber\\
 &\ \ \ +\overline{\widetilde{\chi}_{l}^{0}}(C^{R*}_{lj}P_{L}+C^{L*}_{lj}P_{R})t\widetilde{t}_{j}^{*}
     +\overline{\widetilde{\chi}_{l}^{0}}(D^{R*}_{lk}P_{L}+D^{L*}_{lk}P_{R})b\widetilde{b}_{k}^{*} \,, \nonumber\\
\mathcal{L}_{W\widetilde{\chi}^{0}\widetilde{\chi}^{+}}
 &= W_{\mu}^{-}\overline{\widetilde{\chi}_{l}^{0}}\gamma^{\mu}(E^{L}_{il}P_{L}+E^{R}_{il}P_{R})\widetilde{\chi}_{i}^{+}
     +W_{\mu}^{+}\overline{\widetilde{\chi}_{i}^{+}}\gamma^{\mu}(E^{L*}_{il}P_{L}+E^{R*}_{il}P_{R})\widetilde{\chi}_{l}^{0} \,, \nonumber\\
\mathcal{L}_{W\widetilde{q}\widetilde{q}^{'}}
 &= i(F_{kj}W_{\mu}^{-}\widetilde{b}_{k}^{*}\stackrel{\leftrightarrow}{\partial}_\mu\widetilde{t}_{j}
      +F_{kj}^{*}W_{\mu}^{+}\widetilde{t}_{j}^{*}\stackrel{\leftrightarrow}{\partial}_\mu\widetilde{b}_{k}) \,, \nonumber\\
\mathcal{L}_{Wqq'}
 &= -\frac{g}{\sqrt{2}}(W_{\mu}^{-}\overline{b}\gamma^{\mu}P_{L}t + W_{\mu}^{+}\overline{t}\gamma^{\mu}P_{L}b)\,,
\end{alignat}
where $j,k=(1,2)$, and $P_{L,R}=(1\mp\gamma^{5})/2$\,. The corresponding coupling coefficients are given by
\begin{alignat}{1}
 A^{L}_{ik} &=\frac{gm_{t}}{\sqrt{2}m_{W}s_{\beta}}V^{*}_{i2}R^{\widetilde{b}*}_{k1}\,,\hspace{0.5cm}
   A^{R}_{ik}=\frac{g}{\sqrt{2}}\left[\frac{m_{b}}{m_{W}c_{\beta}}U_{i2}R^{\widetilde{b}*}_{k2}
                                      -\sqrt{2}U_{i1}R^{\widetilde{b}*}_{k1}\right]\,, \nonumber\\
 B^{L}_{ij} &=\frac{gm_{b}}{\sqrt{2}m_{W}c_{\beta}}U^{*}_{i2}R^{\widetilde{t}*}_{j1}\,,\hspace{0.5cm}
   B^{R}_{ij}=\frac{g}{\sqrt{2}}\left[\frac{m_{t}}{m_{W}s_{\beta}}V_{i2}R^{\widetilde{t}*}_{j2}
                                      -\sqrt{2}V_{i1}R^{\widetilde{t}*}_{j1}\right] \,, \nonumber\\
 C^{L}_{lj} &=\frac{g}{\sqrt{2}}\left[-\frac{m_{t}}{m_{W}s_{\beta}}N^{*}_{l4}R^{\widetilde{t}*}_{j1}
                                      +\frac{4}{3}\tan\theta_{W}N^{*}_{l1}R^{\widetilde{t}*}_{j2}\right]\,, \nonumber\\
 C^{R}_{lj} &=\frac{g}{\sqrt{2}}\left[-\frac{m_{t}}{m_{W}s_{\beta}}N_{l4}R^{\widetilde{t}*}_{j2}
              -(N_{l2}+\frac{1}{3}\tan\theta_{W}N_{l1})R^{\widetilde{t}*}_{j1}\right]\,, \nonumber\\
 D^{L}_{lk} &=\frac{g}{\sqrt{2}}\left[-\frac{m_{b}}{m_{W}c_{\beta}}N^{*}_{l3}R^{\widetilde{b}*}_{k1}
                                      -\frac{2}{3}\tan\theta_{W}N^{*}_{l1}R^{\widetilde{b}*}_{k2}\right]\,, \nonumber\\
 D^{R}_{lk} &=\frac{g}{\sqrt{2}}\left[-\frac{m_{b}}{m_{W}c_{\beta}}N_{l3}R^{\widetilde{b}*}_{k2}
              +(N_{l2}-\frac{1}{3}\tan\theta_{W}N_{l1})R^{\widetilde{b}*}_{k1}\right]\,, \nonumber\\
 E^{L}_{il} &=\frac{g}{\sqrt{2}}(\sqrt{2}V^{*}_{i1}N_{l2}-V^{*}_{i2}N_{l4})\,,\hspace{0.5cm}
   E^{R}_{il}=\frac{g}{\sqrt{2}}(\sqrt{2}U_{i1}N^{*}_{l2}+U_{i2}N^{*}_{l3})\,, \nonumber\\
     F_{kj} &=-\frac{g}{\sqrt{2}}R^{\widetilde{b}}_{k1}R^{\widetilde{t}*}_{j1}\,.
\end{alignat}
The total amplitude of the $\widetilde{\chi}_{i}^{\pm}\rightarrow\widetilde{\chi}_{1}^{0}W^{\pm}$ processes can be
written as
 \ba
 \mathcal{M}(\widetilde{\chi}_{i}^{\pm}\rightarrow\widetilde{\chi}_{1}^{0}W^{\pm})
 =\mathcal{M}^{(\pm)}_{Tree}+\mathcal{M}^{(\pm)}_{Loop}
 \ea
with
\begin{alignat}{1}
\mathcal{M}^{(+)}_{Tree} &=\overline{u}_{\widetilde{\chi}_{1}^{0}}(k_{1})\gamma^{\mu}
  (E^{L}_{i1}P_{L}+E^{R}_{i1}P_{R})u_{\widetilde{\chi}_{i}^{+}}(p)\epsilon_{\mu}(k_{2})\,, \nonumber\\
\mathcal{M}^{(-)}_{Tree} &=\overline{v}_{\widetilde{\chi}_{i}^{+}}(-p)\gamma^{\mu}
  (E^{L*}_{i1}P_{L}+E^{R*}_{i1}P_{R})v_{\widetilde{\chi}_{1}^{0}}(-k_{1})\epsilon^{*}_{\mu}(-k_{2})\,, \nonumber\\
\mathcal{M}^{(+)}_{Loop} &=\overline{u}_{\widetilde{\chi}_{1}^{0}}(k_{1})
        \left[(\gamma^{\mu}\Lambda^{L}_{(+)i1}+p^{\mu}\Pi^{L}_{(+)i1})P_{L}\right. \nonumber\\
 &\ \ \ \left.+(\gamma^{\mu}\Lambda^{R}_{(+)i1}+p^{\mu}\Pi^{R}_{(+)i1})P_{R}\right]
        u_{\widetilde{\chi}_{i}^{+}}(p)\epsilon_{\mu}(k_{2})\,, \nonumber\\
\mathcal{M}^{(-)}_{Loop} &=\overline{v}_{\widetilde{\chi}_{i}^{+}}(-p)
        \left[(\gamma^{\mu}\Lambda^{L}_{(-)i1}+p^{\mu}\Pi^{R}_{(-)i1})P_{L}\right. \nonumber\\
 &\ \ \ \left.+(\gamma^{\mu}\Lambda^{R}_{(-)i1}+p^{\mu}\Pi^{L}_{(-)i1})P_{R}\right]
        v_{\widetilde{\chi}_{1}^{0}}(-k_{1})\epsilon_{\mu}^{*}(-k_{2})\,,
\end{alignat}
where $\Lambda^{L,R}_{(\pm)}$ and $\Pi^{L,R}_{(\pm)}$ represent the form factors of vertex corrections contributed
by the one-loop diagrams in Fig.~1, whose expressions are listed in the appendix B. At next-to-leading order, the
\emph{CP} violating asymmetry of Eq.~1 can be obtained as
 \ba
 A_{cp}=\frac{2\rho\mbox{Re}\delta G -12m_{\widetilde{\chi}_{i}^{+}}m_{\widetilde{\chi}_{1}^{0}}m_{W}^{2}\mbox{Re}\delta H
        +\lambda(m_{\widetilde{\chi}_{1}^{0}}\mbox{Re}\delta I+m_{\widetilde{\chi}_{i}^{+}}\mbox{Re}\delta J)}
        {2\rho(|E^{L}_{i1}|^{2}+|E^{R}_{i1}|^{2})-24m_{\widetilde{\chi}_{i}^{+}}m_{\widetilde{\chi}_{1}^{0}}m_{W}^{2}
        \mbox{Re}(E^{L}_{i1}E^{R*}_{i1})}
 \ea
with
\begin{alignat}{1}
   \rho &=m_{W}^{2}(m_{\widetilde{\chi}_{i}^{+}}^{2}+m_{\widetilde{\chi}_{1}^{0}}^{2}-2m_{W}^{2})
          +(m_{\widetilde{\chi}_{i}^{+}}^{2}-m_{\widetilde{\chi}_{1}^{0}}^{2})^{2} \,, \nonumber\\
\lambda &=m_{\widetilde{\chi}_{i}^{+}}^{2}+m_{\widetilde{\chi}_{1}^{0}}^{2}+m_{W}^{2}
           -2m_{\widetilde{\chi}_{i}^{+}}m_{\widetilde{\chi}_{1}^{0}}-2m_{\widetilde{\chi}_{i}^{+}}m_{W}
           -2m_{\widetilde{\chi}_{1}^{0}}m_{W} \,,
\end{alignat}
and
\begin{alignat}{1}
\delta G &=(\Lambda^{L}_{(+)i1}E^{L*}_{i1}+\Lambda^{R}_{(+)i1}E^{R*}_{i1})
           -(\Lambda^{L}_{(-)i1}E^{L}_{i1}+\Lambda^{R}_{(-)i1}E^{R}_{i1}) \,, \nonumber\\
\delta H &=(\Lambda^{L}_{(+)i1}E^{R*}_{i1}+\Lambda^{R}_{(+)i1}E^{L*}_{i1})
           -(\Lambda^{L}_{(-)i1}E^{R}_{i1}+\Lambda^{R}_{(-)i1}E^{L}_{i1}) \,, \nonumber\\
\delta I &=(\Pi^{L}_{(+)i1}E^{L*}_{i1}+\Pi^{R}_{(+)i1}E^{R*}_{i1})
           -(\Pi^{L}_{(-)i1}E^{L}_{i1}+\Pi^{R}_{(-)i1}E^{R}_{i1}) \,, \nonumber\\
\delta J &=(\Pi^{L}_{(+)i1}E^{R*}_{i1}+\Pi^{R}_{(+)i1}E^{L*}_{i1})
           -(\Pi^{L}_{(-)i1}E^{R}_{i1}+\Pi^{R}_{(-)i1}E^{L}_{i1}) \,.
\end{alignat}
From Eq.~(6)-(8), it can directly be seen that there is no \emph{CP} violation at tree level because of all of the
form factors vanishing. However, for generating \emph{CP} violation loop corrections, \emph{viz}. nonvanishing
form factors, and complex couplings, \emph{viz}. nonvanishing \emph{CP} violating phases, are essential factors.

\begin{center}
\textbf{III. NUMERICAL RESULTS}
\end{center}

In this section, we will illustrate numerical results of the \emph{CP} violating asymmetry based on the MSSM
parameter space at the electroweak scale allowed by present data constrains \cite{11}. In order not to vary too
many parameters, we assume the grand unified theory relation for the gaugino mass parameters, $M_{1}\approx
0.5M_{2}$\,. In this case, the phase of the gaugino sector can be rotated away. In addition, since the phase of
Higgs mixing parameter, $\varphi_{\mu}$\,, is highly constrained by the EDMs of electron and neutron \cite{4}, we
take $\varphi_{\mu}=0$\,. For the mass parameters of the squark sectors, we fix simply the relations:
$M_{\widetilde{U}}:M_{\widetilde{Q}}:M_{\widetilde{D}}\approx 0.85:1:1.05$\,. Thus, in our following numerical
analyses the input parameters contain only:
$M_{2},M_{\widetilde{Q}},|\mu|,|A_{t}|,|A_{b}|,\varphi_{t},\varphi_{b}$ and $\tan\beta$. Corresponding to a set of
representative values in the parameter space as follows:
\begin{alignat}{1}
M_{2}=250\ \mbox{GeV},\hspace{0.5cm} M_{\widetilde{Q}}=450\ \mbox{GeV},\hspace{0.5cm} |\mu|=500\ \mbox{GeV}, \nonumber\\
|A_{t}|=|A_{b}|=500\ \mbox{GeV},\hspace{0.3cm}\varphi_{t}=\varphi_{b}=\frac{\pi}{4}, \hspace{0.3cm}\tan\beta=5\
\mbox{or}\ 40\,,
\end{alignat}
Table\,1 list explicitly the relevant sparticle masses. It can be seen that because of the large Yukawa couplings
of the third generation squarks, the mixing between stops or sbottoms can be very strong, especially for the case
of the high value of $\tan\beta=40$\,. Moreover, the masses of $\widetilde{\chi}_{1}^{0}$ and
$\widetilde{\chi}_{1}^{+}$ approximate to the values of $M_{1}$ and $M_{2}$\,, respectively, while the mass of
$\widetilde{\chi}_{2}^{+}$ is about the value of $|\mu|$\,. In fact, this is generic in the region $|\mu|\geqslant
M_{2}$ \cite{9}. Therefore, if only $M_{2}$ and $|\mu|$ are not too low, both
$\widetilde{\chi}_{1}^{\pm}\rightarrow\widetilde{\chi}_{1}^{0}W^{\pm}$ and
$\widetilde{\chi}_{2}^{\pm}\rightarrow\widetilde{\chi}_{1}^{0}W^{\pm}$ are kinematically allowed.
\begin{table}
\centering
\begin{tabular}{c|ccccccc}
\hline
 $\tan\beta$ & $m_{\widetilde{\chi}_{1}^{+}}$ & $m_{\widetilde{\chi}_{2}^{+}}$ & $m_{\widetilde{\chi}_{1}^{0}}$
 & $m_{\widetilde{t}_{1}}$ & $m_{\widetilde{t}_{2}}$ & $m_{\widetilde{b}_{1}}$ & $m_{\widetilde{b}_{2}}$ \\
\hline
  5 & 236 & 519 & 122 & 350 & 532 & 449 & 477 \\
 40 & 241 & 517 & 124 & 336 & 541 & 361 & 548 \\
\hline
\end{tabular}
\caption{The relevant sparticle masses (in GeV) for parameter sets in Eq.~(9).}
\end{table}

We now consider only the $\widetilde{\chi}_{2}^{\pm}\rightarrow\widetilde{\chi}_{1}^{0}W^{\pm}$ channel, and
demonstrate in turn the dependence of its \emph{CP} violating asymmetry $A_{cp}$ on various choices of the
parameters. Because our results are not sensitive to $M_{2}$\,, it will be fixed. In Fig.~2\,, Absolute value of
the \emph{CP} asymmetry, $|A_{cp}|$\,, is shown as a function of the Higgs mixing parameter $|\mu|$\,. The four
curves in this figure are corresponding to four combined choices of $\tan\beta=5$ (or 40) and
$\varphi_{t}=\varphi_{b}=\pi/4$ (or $\pi/2$), respectively. The other parameters are fixed by Eq.~(9). From these
curves, for instance, for the short-dashed line of $\tan\beta=5$ and $\varphi_{t}=\varphi_{b}=\pi/4$\,, one can
distinguish the thresholds of $b\widetilde{t}_{1}$ at $|\mu|\approx 295$ GeV, $b\widetilde{t}_{2}$ at
$|\mu|\approx 495$ GeV, $t\widetilde{b}_{1}$ at $|\mu|\approx 610$ GeV, $t\widetilde{b}_{2}$ at $|\mu|\approx 650$
GeV. In the region of small $|\mu|$\,, the contributions to $|A_{cp}|$ come from the stop-bottom-sbottom loop of
Fig.~2~(b) and the stop-bottom-top loop of Fig.~2~(d). As can be seen, once
$m_{\widetilde{\chi}_{2}^{+}}>m_{b}+m_{\widetilde{t}_{1}}$\,, $|A_{cp}|$ can sharply go up to order of $10^{-3}$
($10^{-2}$) for $\tan\beta=5$ ($\tan\beta=40$)\,. As the value of $|\mu|$ increasing, the
$\widetilde{\chi}_{2}^{+}\rightarrow b\widetilde{t}_{2}$ channel is open, $|A_{cp}|$ can further rise to several
per cent for whether $\tan\beta=5$ or $\tan\beta=40$. For larger $|\mu|$\,, it should however be taken into
account that the contributions of diagram (a) and (c) in Fig.~2 though they are relatively smaller.
\begin{figure}
\centering
\includegraphics[totalheight=5.5cm]{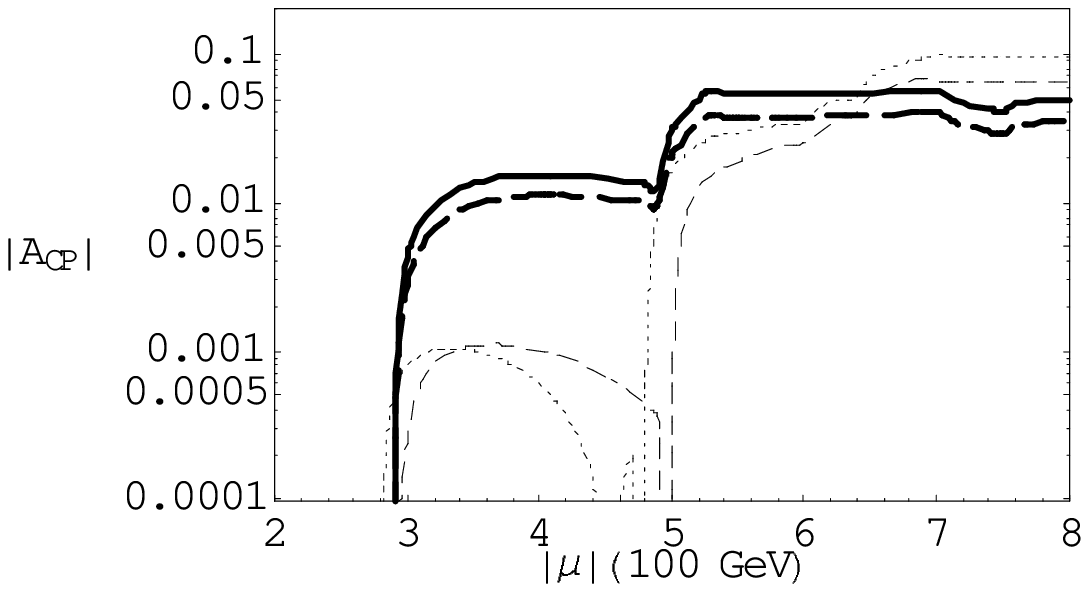}
\caption{Absolute value of $A_{cp}$ as a function of $|\mu|$. The thinner short-dashed line (dotted line) is for
$\tan\beta=5,\varphi_{t}=\varphi_{b}=\pi/4\ (\pi/2)$, while the thicker long-dashed line (solid line) is for
$\tan\beta=40,\varphi_{t}=\varphi_{b}=\pi/4\ (\pi/2)$, respectively. The other parameters are fixed by Eq.~(9).}
\end{figure}

\begin{figure}
\centering
\includegraphics[totalheight=5.5cm]{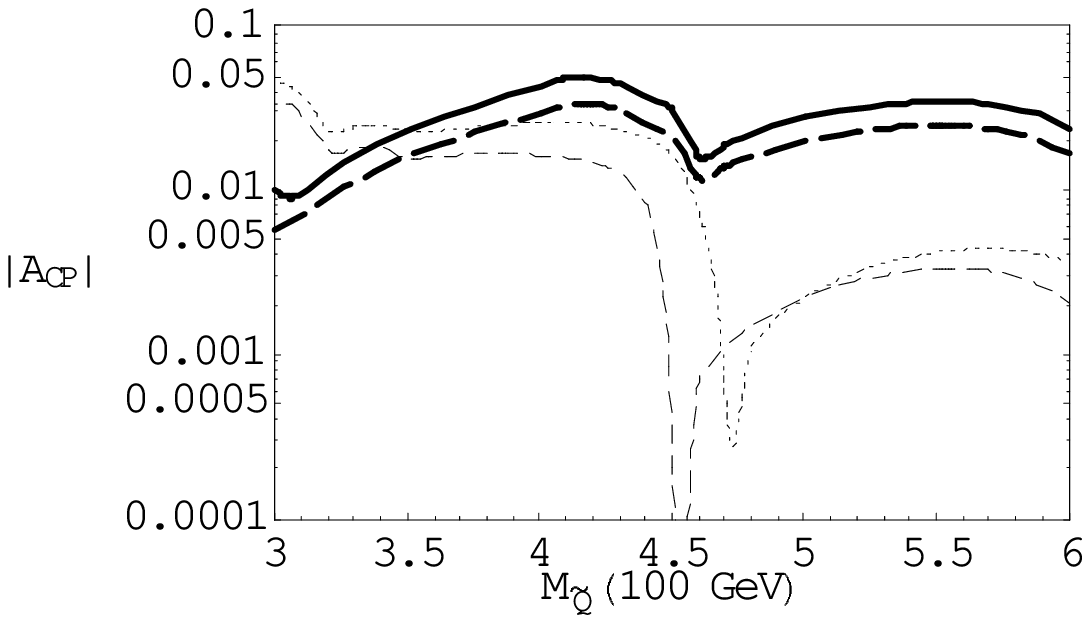}
\caption{$|A_{cp}|$ as a function of $M_{\widetilde{Q}}$\,. The thinner short-dashed line (dotted line) is for
$\tan\beta=5,\varphi_{t}=\varphi_{b}=\pi/4\ (\pi/2)$, while the thicker long-dashed line (solid line) is for
$\tan\beta=40,\varphi_{t}=\varphi_{b}=\pi/4\ (\pi/2)$, respectively. The other parameters are fixed by Eq.~(9).}
\end{figure}
In Fig.~3\,, we plot $|A_{cp}|$ as a function of the left-handed soft-SUSY-breaking squark mass
$M_{\widetilde{Q}}$ for the same arrangement of $\tan\beta$ and $\varphi_{t},\varphi_{b}$ as in Fig.~2\,. The
other parameters are still given by Eq.~(9). Here it is more clearly visible that the threshold of
$b\widetilde{t}_{2}$ is at $M_{\widetilde{Q}}\approx 455$ (470) GeV for $\tan\beta=5$ and
$\varphi_{t}=\varphi_{b}=\pi/4$ ($\pi/2$)\,, and at $M_{\widetilde{Q}}\approx 460$ (465) GeV for $\tan\beta=40$
and $\varphi_{t}=\varphi_{b}=\pi/4$ ($\pi/2$)\,. For $\tan\beta=40$\,, $|A_{cp}|$ is essentially of the order of
$10^{-2}$ in the region $M_{\widetilde{Q}}=300\sim600$ GeV. For $\tan\beta=5$\,, in the lighter squark region,
namely above the threshold of $\widetilde{\chi}_{2}^{+}\rightarrow b\widetilde{t}_{2}$\,, $|A_{cp}|$ is also of
the order of $10^{-2}$; in the heavier squark region, the $\widetilde{\chi}_{2}^{+}\rightarrow b\widetilde{t}_{2}$
channel is close, $|A_{cp}|$ lower to order of $10^{-3}$\,.

\begin{figure}
\centering
\includegraphics[totalheight=5.5cm]{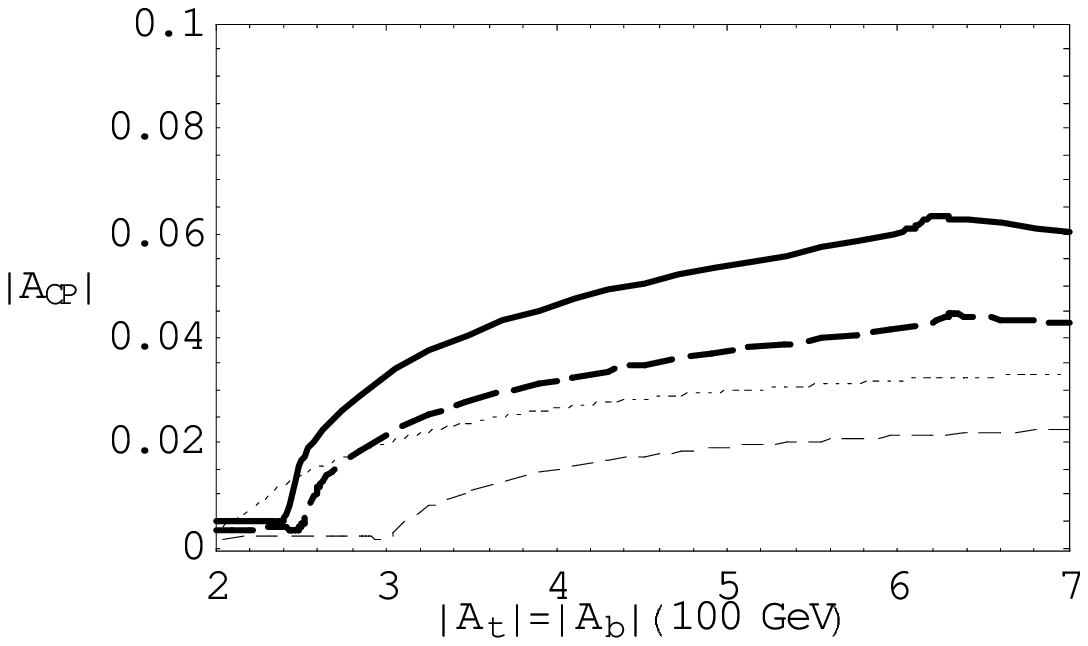}
\caption{$|A_{cp}|$ as a function of $|A_{t}|=|A_{b}|$ for $|\mu|=550$ GeV. The thinner short-dashed line (dotted
line) is for $\tan\beta=5,\varphi_{t}=\varphi_{b}=\pi/4\ (\pi/2)$, while the thicker long-dashed line (solid line)
is for $\tan\beta=40,\varphi_{t}=\varphi_{b}=\pi/4\ (\pi/2)$, respectively. The other parameters are fixed by
Eq.~(9).}
\end{figure}
The \emph{CP} asymmetry as a function of the trilinear coupling $|A_{t}|=|A_{b}|$ is illustrated in Fig.~4\,. Here
we take $|\mu|=550$ GeV, the other parameters are still fixed by Eq.~(9) and the choices of
$\tan\beta,\varphi_{t},\varphi_{b}$ are as same as the previous. For the smaller values of $|A_{t}|=|A_{b}|$\,,
$|A_{cp}|$ is very small and below $\mathcal{O}(0.01)$\,. As $|A_{t}|=|A_{b}|$ increase from small to large,
$|A_{cp}|$ rise to a few per cent. On the other hand, changing $\tan\beta=5$ to $\tan\beta=40$\,, the values of
$|A_{cp}|$ are raised two per cent or so; varying $\varphi_{t}=\varphi_{b}$ from $\pi/4$ to $\pi/2$\,, $|A_{cp}|$
is enhanced about 0.01\,.

\begin{figure}
\centering
\includegraphics[totalheight=5.5cm]{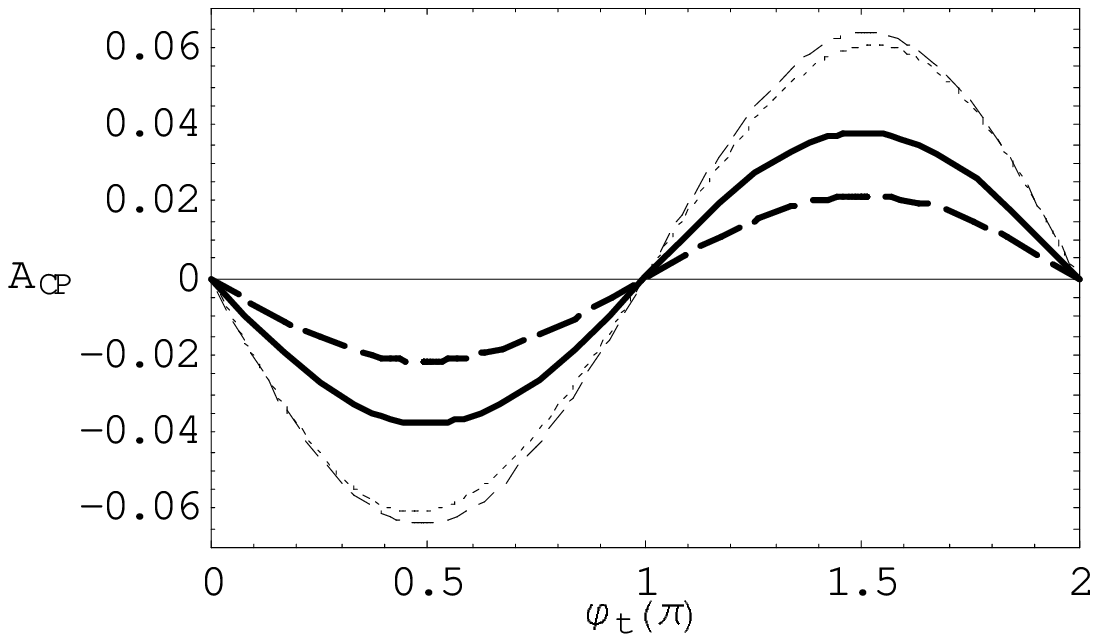}
\caption{$A_{cp}$ as a function of $\varphi_{t}$ for $M_{\widetilde{Q}}=400$ GeV and $|\mu|=600$ GeV. The thinner
short-dashed line (dotted line) is for $\tan\beta=5,\varphi_{b}=0\ (\varphi_{t})$, while the thicker long-dashed
line (solid line) is for $\tan\beta=40,\varphi_{b}=0\ (\varphi_{t})$, respectively. The other parameters are fixed
by Eq.~(9).}
\end{figure}
Fig.~5 shows the dependence of $A_{cp}$ on the \emph{CP} violating phase $\varphi_{t}$ for the four combined
choices of $\tan\beta=5/40$ and $\varphi_{b}=0/\varphi_{t}$\,. Here we take $|\mu|=600$ GeV and
$M_{\widetilde{Q}}=400$ GeV, the other parameters are given by Eq.~(9)\,. As expected, $A_{cp}$ shows a $\sim$
$\sin\varphi_{t}$ dependence. For the case of $\tan\beta=5$\,, because the value of $|\mu|$ is relatively larger
to the value of $M_{\widetilde{Q}}$\,, all of the $b\widetilde{t}_{1,2}$\,, $t\widetilde{b}_{1,2}$ channels are
open. At the two maximal stop phases, $\varphi_{t}=\pi/2,\ 3\pi/2$\,, the \emph{CP} asymmetry can reach 6\%\,. In
this case, because the mixing between sbottoms is not so large, the influence of the phase $\varphi_{b}$ is
relatively smaller. For the case of $\tan\beta=40$\,, however, the mixing between sbottoms is comparable with the
mixing between stops, $\varphi_{b}$ can obviously change $A_{cp}$ by up to 40\% at $\varphi_{t}=\pi/2,\ 3\pi/2$\,.

\begin{center}
\textbf{IV. CONCLUSIONS}
\end{center}

In this work, we have discuss the possible \emph{CP} violation in the decays of charginos and neutralinos in the
MSSM. For the decay of the heaviest chargino into the lightest neutralino and a $W$ boson, we have explicitly
presented analytical and numerical results of the \emph{CP} violating asymmetry. The decay rate asymmetry
originates from the large \emph{CP} violating phases and one-loop corrections of the third generation squark
sectors. The \emph{CP} asymmetry can typically reach several per cent, depending mainly on the phases and
$\tan\beta$\,. For a larger mixing between the stops (sbottoms), the \emph{CP} asymmetry is in particular evident.
At a future $e^{+}e^{-}$ linear collider with $\sqrt{s}=500\sim 1000$ GeV \cite{12a}, the cross section of
chargino pair production is around $\mathcal{O}(1)$ pb \cite{12b}. With a luminosity $\mathcal{L}=100$
$\mbox{fb}^{-1}$, about several hundred signal events can be expected since the decay channels has a larger
branching ratio and a cleaner final signal. Therefore, Analyzing these final states would allow to measure the
important couplings of squark sectors and determine the \emph{CP} violating phases. Thus providing a opportunity
for detecting directly the \emph{CP} violating effects of the MSSM.

\begin{center}
\textbf{ACKNOWLEDGMENTS}
\end{center}

One of the authors, W. M. Yang, thanks M. Z. Yang for helpful discussions. This work is in part supported by
National Natural Science Foundation of China.

\begin{center}
\textbf{APPENDIX A: SUSY PARTICLES MASSES AND MIXING}
\end{center}

To fix our notation, we simply summarize in this appendix the masses and mixing of the relevant sparticles. The
mass matrices of charginos and neutralinos are respectively given by
 \ba M_C=\left(\begin{array}{cc} M_2 & \sqrt{2}m_W s_\beta \\ \sqrt{2}m_W c_\beta & \mu\end{array}\right) \ea
and
 \ba M_N=\left(\begin{array}{cccc}
     M_1 & 0 & -m_Z s_W c_\beta & m_Z s_W s_\beta \\
     0 & M_2 & m_Z c_W c_\beta & -m_Z c_W s_\beta \\
     -m_Z s_W c_\beta & m_Z c_W c_\beta & 0 & -\mu\\
     m_Z s_W s_\beta & -m_Z c_W s_\beta & -\mu & 0
  \end{array}\right)\,, \ea
where we have used the abbreviations: $s_W=\sin\theta_W$, $c_\beta=\cos\beta$, etc. $M_1$ and $M_2$ are the
soft-SUSY-breaking gaugino mass parameters. $\mu$ is the Higgs mixing parameter. $\tan\beta$ is the ratio of the
two Higgs vacuum expectation values. $\mu$ and one of $M_i\ (i=1,2)$ can be complex. The chargino mass matrix can
be diagonalized by two unitary matrices $U$ and $V$\,,
 \ba U^{*} M_C V^{\dagger}=\mbox{diag}\,(m_{\widetilde{\chi}^{+}_1}\,,\,m_{\widetilde{\chi}^{+}_2})\,, \ea
where $m_{\widetilde{\chi}^{+}_{1,2}}$ are the masses of the physical chargino states. The neutralino mass matrix
can be diagonalized by a single unitary matrix $N$\,,
 \ba N^{*} M_N N^{\dagger}=\mbox{diag}\,(m_{\widetilde{\chi}^{0}_1}\,,\,m_{\widetilde{\chi}^{0}_2}\,,
                     \,m_{\widetilde{\chi}^{0}_3}\,,\,m_{\widetilde{\chi}^{0}_4})\,, \ea
where $m_{\widetilde{\chi}^{0}_{l}}\ (l=1-4)$ are the masses of the physical neutralino states. The expressions of
mass eigenvalues and mixing matrix elements for chargions and neutralinos can be found in Ref \cite{9}.

The mass-squared matrices of stops and sbottoms in the left-right basis can be written as
 \ba M^2_{\widetilde{q}}=\left(\begin{array}{cc}
     m^2_{LL} & m^2_{LR} \\ m^{2*}_{LR} & m^2_{RR} \end{array}\right) \hspace{0.5cm} (q=t/b) \ea
 with
\begin{alignat}{1}
 m^2_{LL} &=M^2_{\widetilde{Q}}+ m^2_q + m^2_Z\cos2\beta(T^{q}_{3}-Q_{q}s^{2}_{W})\,, \nonumber\\
 m^2_{RR} &=M^2_{\widetilde{U}/\widetilde{D}}+ m^2_q + m^2_Z\cos2\beta Q_{q}s^{2}_{W}\,, \nonumber\\
 m^2_{LR} &=m_q(A^{*}_q - \mu\cot\beta)/m_q(A^{*}_q - \mu\tan\beta)\,,
\end{alignat}
where $M_{\widetilde{Q}}$ and $M_{\widetilde{U}}$ ($M_{\widetilde{D}}$) are the left- and right-handed
soft-SUSY-breaking stop (sbottom) masses, respectively. The soft-SUSY-breaking trilinear couplings $A_q$ are
complex parameters,
 \ba A_t=|A_t|e^{i\,\varphi_{t}}\,, \hspace{0.6cm} A_b=|A_b|e^{i\,\varphi_{b}} \ea
with \emph{CP} violating phases $\varphi_{t}$ and $\varphi_{b}$\,. The mass-squared matrix $M^2_{\widetilde{q}}$
can be diagonalized by a unitary matrix $R^{\widetilde{q}}$\,,
 \ba R^{\widetilde{q}}\, M^2_{\widetilde{q}}\, R^{\widetilde{q}\,\dagger}
     =\mbox{diag}\,(m^2_{\widetilde{q}_1}\,,\,m^2_{\widetilde{q}_2})\,, \ea
 where the diagonalization matrix can be parameterized as
 \ba R^{\widetilde{q}}=\left(\begin{array}{cc}\cos\theta_q & \sin\theta_q e^{i\,\delta_q} \\
 -\sin\theta_q e^{-i\,\delta_q} & \cos\theta_q\end{array}\right) \ea
 with
 \ba \delta_t=\arg\,(A^*_t-\mu\cot\beta)\,, \hspace{0.3cm} \delta_b=\arg\,(A^*_b-\mu\tan\beta)\,.\ea
The squark mass eigenvalues and mixing angle are then given as\\
\begin{alignat}{1}
m^2_{\widetilde{q}_{1,2}}
   &=\frac{1}{2}\left[m^2_{LL}+m^2_{RR}\mp\sqrt{(m^2_{LL}-m^2_{RR})^2+4|m^2_{LR}|^2}\right]\,, \nonumber\\
\tan\theta_{q} &=\frac{2\,|m^2_{LR}|}{m^2_{LL}-m^2_{RR}+m^2_{\widetilde{q}_1}-m^2_{\widetilde{q}_2}}\ .
\end{alignat}

\begin{center}
\textbf{APPENDIX B: THE RELEVANT FORM FACTORS}
\end{center}

In this appendix, we list the form factors of the one-loop corrections in Fig.~1. We give only the form factors of
the decay $\widetilde{\chi}_{i}^{+}\rightarrow\widetilde{\chi}_{l}^{0}W^{+}$\,, $\Lambda^{L,R}_{(+)il}$ and
$\Pi^{L,R}_{(+)il}$\,. For that of the decay $\widetilde{\chi}_{i}^{-}\rightarrow\widetilde{\chi}_{l}^{0}W^{-}$\,,
$\Lambda^{L,R}_{(-)il}$ and $\Pi^{L,R}_{(-)il}$\,, can correspondingly be obtained by conjugating all the
couplings in $\Lambda^{L,R}_{(+)il}$ and $\Pi^{L,R}_{(+)il}$\,. Corresponding to the four one-loop diagram
contributions, the form factors $\Lambda^{L,R}_{(+)il}$ and $\Pi^{L,R}_{(+)il}$ (in the following we omit ``(+)")
can be divided into four parts,
\begin{alignat}{1}
\Lambda^{L,R}_{il} &=\Lambda^{(1)L,R}_{il}+\Lambda^{(2)L,R}_{il}+\Lambda^{(3)L,R}_{il}+\Lambda^{(4)L,R}_{il}\,,\nonumber\\
    \Pi^{L,R}_{il} &=\Pi^{(1)L,R}_{il}+\Pi^{(2)L,R}_{il}+\Pi^{(3)L,R}_{il}+\Pi^{(4)L,R}_{il}
\end{alignat}
with
\begin{alignat*}{1}
\Lambda^{(1)L}_{il} &=\frac{3}{8\pi^{2}}\sum_{j,k}A^{L}_{ik}C^{L*}_{lj}F_{kj}C^{(1)}_{24}\,, \ \
\Lambda^{(1)R}_{il} =\Lambda^{(1)L}_{il}\ (L\leftrightarrow R)\,, \\
\Lambda^{(2)L}_{il} &=\frac{3}{8\pi^{2}}\sum_{j,k}B^{L*}_{ij}D^{L}_{lk}F_{kj}C^{(2)}_{24}\,, \ \
\Lambda^{(2)R}_{il} =\Lambda^{(2)L}_{il}\ (L\leftrightarrow R)\,, \\
\Lambda^{(3)L}_{il}
&=\frac{3g}{16\sqrt{2}\pi^{2}}\sum_{k}\left\{m_{b}m_{\widetilde{\chi}_{l}^{0}}A^{L}_{ik}D^{R*}_{lk}C^{(3)}_{12}
  -m_{t}m_{\widetilde{\chi}_{i}^{+}}A^{R}_{ik}D^{L*}_{lk}(C^{(3)}_{0}+C^{(3)}_{11})\right. \\
&\ \ \ \left.+A^{L}_{ik}D^{L*}_{lk}\left[\frac{1}{2}+2C^{(3)}_{24}+m^{2}_{\widetilde{\chi}_{i}^{+}}
 (C^{(3)}_{11} -C^{(3)}_{12}+C^{(3)}_{21}-C^{(3)}_{23})\right.\right. \\
&\ \ \ \left.\left.+m^{2}_{\widetilde{\chi}_{l}^{0}}(C^{(3)}_{22}-C^{(3)}_{23})
  +m^{2}_{W}(C^{(3)}_{12}+C^{(3)}_{23})\right]\right\}\,,
\end{alignat*}
\begin{alignat}{1}
\Lambda^{(3)R}_{il} &=\frac{3g}{16\sqrt{2}\pi^{2}}\sum_{k}\left\{-m_{t}m_{b}A^{R}_{ik}D^{R*}_{lk}C^{(3)}_{0}
+m_{b}m_{\widetilde{\chi}_{i}^{+}}A^{L}_{ik}D^{R*}_{lk}C^{(3)}_{11}\right. \nonumber\\
&\ \ \ \left.-m_{t}m_{\widetilde{\chi}_{l}^{0}}A^{R}_{ik}D^{L*}_{lk}(C^{(3)}_{0}+C^{(3)}_{12})
+m_{\widetilde{\chi}_{i}^{+}}m_{\widetilde{\chi}_{l}^{0}}A^{L}_{ik}D^{L*}_{lk}(C^{(3)}_{11}-C^{(3)}_{12})\right\}\,, \nonumber\\
\Lambda^{(4)L,R}_{il} &=\Lambda^{(3)L,R}_{il}\ \left(A^{L,R}_{ik}\leftrightarrow C^{L,R}_{lk}\,,
\ D^{L,R}_{lk}\leftrightarrow B^{L,R}_{ik}\,,\ m_{\widetilde{\chi}_{i}^{+}}\leftrightarrow m_{\widetilde{\chi}_{l}^{0}}\,,\right. \nonumber\\
&\ \ \ \left.C^{(3)}_{0,11,12,21,22,23,24}\leftrightarrow C^{(4)}_{0,11,12,21,22,23,24}\right)\,, \nonumber\\
\Pi^{(1)L}_{il} &=-\frac{3}{16\pi^{2}}\sum_{j,k}F_{kj}\left[m_{t}A^{L}_{ik}C^{R*}_{lj}(C^{(1)}_{11}-C^{(1)}_{12})
+m_{\widetilde{\chi}_{l}^{0}}A^{L}_{ik}C^{L*}_{lj}(C^{(1)}_{22}-C^{(1)}_{23})\right. \nonumber\\
&\ \ \ \left.+m_{\widetilde{\chi}_{i}^{+}}A^{R}_{ik}C^{R*}_{lj}(C^{(1)}_{11}
     -C^{(1)}_{12}+C^{(1)}_{21}-C^{(1)}_{23})\right]\,, \nonumber\\
\Pi^{(1)R}_{il} &=\Pi^{(1)L}_{il}\ (L\leftrightarrow R)\,, \nonumber\\
\Pi^{(2)L}_{il} &=-\frac{3}{16\pi^{2}}\sum_{j,k}F_{kj}\left[m_{b}B^{R*}_{ij}D^{L}_{lk}(C^{(2)}_{11}-C^{(2)}_{12})
+m_{\widetilde{\chi}_{i}^{+}}B^{L*}_{ij}D^{L}_{lk}(C^{(2)}_{22}-C^{(2)}_{23})\right. \nonumber\\
&\ \ \ \left.+m_{\widetilde{\chi}_{l}^{0}}B^{R*}_{ij}D^{R}_{lk}(C^{(2)}_{11}
     -C^{(2)}_{12}+C^{(2)}_{21}-C^{(2)}_{23})\right]\,, \nonumber\\
\Pi^{(2)R}_{il} &=\Pi^{(2)L}_{il}\ (L\leftrightarrow R)\,, \nonumber\\
\Pi^{(3)L}_{il} &=\frac{3g}{16\sqrt{2}\pi^{2}}\sum_{k}\left[m_{b}A^{L}_{ik}D^{R*}_{lk}C^{(3)}_{12}
+m_{\widetilde{\chi}_{l}^{0}}A^{L}_{ik}D^{L*}_{lk}(C^{(3)}_{22}-C^{(3)}_{23})\right]\,, \nonumber\\
\Pi^{(3)R}_{il} &=\frac{3g}{16\sqrt{2}\pi^{2}}\sum_{k}\left[-m_{t}A^{R}_{ik}D^{L*}_{lk}(C^{(3)}_{0}+C^{(3)}_{11})\right. \nonumber\\
&\ \ \ \left.+m_{\widetilde{\chi}_{i}^{+}}A^{L}_{ik}D^{L*}_{lk}(C^{(3)}_{11}-C^{(3)}_{12}+C^{(3)}_{21}-C^{(3)}_{23})\right]\,, \nonumber\\
\Pi^{(4)L,R}_{il} &=\Pi^{(3)L,R}_{il}\ \left(A^{L,R}_{ik}\leftrightarrow C^{L,R}_{lk}\,,
\ D^{L,R}_{lk}\leftrightarrow B^{L,R}_{ik}\,,\ m_{\widetilde{\chi}_{i}^{+}}\leftrightarrow m_{\widetilde{\chi}_{l}^{0}}\,,\right. \nonumber\\
&\ \ \ \left.C^{(3)}_{0,11,12,21,22,23}\leftrightarrow C^{(4)}_{0,11,12,21,22,23}\right)\,,
\end{alignat}
where
\begin{alignat}{1}
C^{(1)}_{11,12,21,22,23,24} &=
C_{11,12,21,22,23,24}(m^2_{\widetilde{\chi}_{l}^{0}}\,,\,m^2_{\widetilde{\chi}_{i}^{+}}\,,\,m^2_W\,,\,
                        m^2_{\widetilde{t}_{j}}\,,\,m^2_t\,,\,m^2_{\widetilde{b}_{k}})\,, \nonumber\\
C^{(2)}_{11,12,21,22,23,24} &=
C_{11,12,21,22,23,24}(m^2_{\widetilde{\chi}_{l}^{0}}\,,\,m^2_{\widetilde{\chi}_{i}^{+}}\,,\,m^2_W\,,\,
                        m^2_{\widetilde{b}_{k}}\,,\,m^2_b\,,\,m^2_{\widetilde{t}_{j}})\,, \nonumber\\
C^{(3)}_{0,11,12,21,22,23,24} &=
C_{0,11,12,21,22,23,24}(m^2_{\widetilde{\chi}_{l}^{0}}\,,\,m^2_{\widetilde{\chi}_{i}^{+}}\,,\,m^2_W\,,\,
                        m^2_b\,,\,m^2_{\widetilde{b}_{k}}\,,\,m^2_t)\,, \nonumber\\
C^{(4)}_{0,11,12,21,22,23,24} &=
C_{0,11,12,21,22,23,24}(m^2_{\widetilde{\chi}_{l}^{0}}\,,\,m^2_{\widetilde{\chi}_{i}^{+}}\,,\,m^2_W\,,\,
                        m^2_t\,,\,m^2_{\widetilde{t}_{j}}\,,\,m^2_b)\,.
\end{alignat}
The definitions and numerical calculation formulae of the Passarino-Veltman  one-, two-, and three-point functions
 are adopted from Refs \cite{13}.

\end{document}